\documentclass{article}
\usepackage{mlspconf,amsmath,graphicx}
\usepackage{verbatim}
\usepackage{amsmath}
\usepackage{amssymb}        
\usepackage{xcolor}
\usepackage{empheq,float}
\usepackage{graphicx,caption,subcaption,balance}
\copyrightnotice{978-1-7281-6662-9/20/\$31.00 {\copyright}2020 IEEE}

\toappear{2020 IEEE International Workshop on Machine Learning for Signal Processing, Sept.\ 21--24, 2020, Espoo, Finland}

\input{mysymbol.sty}
\usepackage{algorithm}
\usepackage{algpseudocode}
\usepackage{mathrsfs}
\usepackage{psfrag}

\title{Topology-Aware Joint Graph Filter and Edge Weight
Identification for Network Processes}

\name{Alberto Natali, Mario Coutino and Geert Leus}
\address{Faculty of Electrical Engineering, Mathematics and Computer Science\\ Delft University of Technology, Delft, The Netherlands  \\
E-mails: \{a.natali; m.a.coutinominguez; g.j.t.leus\}@tudelft.nl}
\begin{document}
\ninept
\maketitle
\begin{abstract} 
Data defined over a network have been successfully modelled by means of graph filters. However, although in many scenarios the connectivity of the network is known, e.g., smart grids, social networks, etc., the lack of well-defined interaction weights hinders the ability to model the observed networked data using graph filters. Therefore, in this paper, we focus on the joint identification of coefficients and graph weights defining the graph filter that best models the observed input/output network data. 

While these two problems have been mostly addressed separately, we here propose an iterative method that exploits the knowledge of the support of the graph for the joint identification of graph filter coefficients and edge weights.
We further show that our iterative scheme guarantees a non-increasing cost at every iteration, ensuring a globally-convergent behavior. Numerical experiments confirm the applicability of our proposed approach. 
\end{abstract}
\vspace{-3mm}

\section{Introduction}\vspace{-2mm}
\label{sec:intro}

The increasing amount of \textit{networked} data, also conceptualized as graph signals within the graph signal processing (GSP) field \cite{shuman2013emerging} \cite{6409473}, has gained a lot of attention in the scientific community. Due to this, many signal processing tasks have been adapted towards their networked counterpart, as extensively detailed in \cite{ortega2018graph}. 

In the graph setting, it is common to parameterize network processes through graph filters, due to their versatility and their natural distributed implementation \cite{shuman2018distributed} \cite{segarra2017optimal}. They play an important role within GSP, with applications ranging from reconstruction \cite{narang2013signal} \cite{girault2014semi} \cite{Isufi2018DistributedWR}, denoising \cite{onuki2016graph} and classification \cite{7472541}, to forecasting \cite{isufi2019forecasting} \cite{natali2020filter} and  (graph-)convolutional neural networks \cite{gama2018convolutional}. Notable recent advances in such structures are \cite{coutino2019advances}, which generalizes state-of-the-art graph filters to filters where every node weights the signal of its neighbors with different values, and \cite{segarra2016blind}, which extends the classical problem of blind system identification or blind de-convolution to the graph setting.

Given the structure of the graph, encoded by the so-called graph shift operator (GSO) \cite{6409473}, and assuming a process modelled by a graph filter, identifying an underlying network process from input/output networked data amounts to estimate the graph filter coefficients, thus alleviating the estimation workload \cite{6409473} \cite{liu2018filter}. A key assumption in graph filtering is the \textit{knowledge} of the GSO, which can be obtained from some other field of research or can be estimated from historical data. The latter relates to network topology inference or graph learning which, in recent years, has experienced an exponentially-increasing scientific interest, see, e.g.,~\cite{mateos2019connecting} \cite{dong2019learning} \cite{giannakis2018topology}.

Related to the scenario we are going to consider, there are also works that model the observed signal as the output of an unknown graph filter over an unknown graph. In \cite{segarra2017network}, a two-step GSO identification approach is taken, where first the GSO's eigenvectors are identified from the diffused (stationary) graph signals and then the GSO's eigenvalues are estimated based on some general properties of the GSO. In \cite{shafipour2018identifying}, the work of \cite{segarra2017network} is extended to non-stationary graph signals, entailing the solution of a system of quadratic matrix equations. Using the same approach, the problem of directed network topology identification is investigated in \cite{shapifour}. Note, though, that none of these above works focuses on estimating the related graph filter. More similar to our work, is the approach of \cite{7763882}, where not only the GSO but also the filter taps are learned. Although the context of \cite{7763882} is different, in that work, a general linear filter operator is estimated from the data and then both the GSO and the filter taps are estimated from it.

All the previous approaches rely on  a multi-step algorithm and only exploit some general properties of the GSO, e.g., sparsity. In addition, in many practical networks such as social and supply networks, the support of the graph is a priori known, that is, the connections between different entities of the network are already known, yet their importance might be unknown. And this information is not directly handled by the above algorithms.

Motivated by the above reasons, this work aims to jointly estimate the graph filter coefficients and the weights of the network topology. This joint approach leads to an optimization problem that is non-convex. We tackle the non-convexity of the problem by  building on sequential convex programming (SCP), a local optimization tool for non-convex problems that leverages the
convex optimization machinery. We show that an alternating minimization between the filter coefficients and the GSO guarantees that the objective function value at each iteration is non-increasing, obtaining a globally convergent method. 

\section{Preliminaries}
In this section, we introduce the GSP background material necessary for the rest of the paper, including the formal definition of graph signals and the core concepts of graph filtering and topology identification.

\textbf{Graph Signal Processing} We consider the case in which the data of interest live in a non-Euclidean domain, described by the undirected graph $\ccalG=\left(\ccalV,\ccalE, \bbS \right)$,  where $\ccalV=\{1, \ldots, N\}$ is the set of nodes (or vertices),  $\ccalE \subseteq \mathcal{V} \times \mathcal{V}$ is the set of edges, and $\bbS$ is a symmetric $N \times N$ matrix that represents the graph structure. The matrix $\bbS$ is called the graph shift operator (GSO) \cite{6409473}, whose entries $\left[\bbS\right]_{ij}$ for $i\neq j$ are different from zero only if nodes $i$ and $j$ are connected by an edge. Typical choices of the GSO include the (weighted) adjacency matrix $\bbW$ \cite{6409473} and the graph Laplacian $\bbL$ \cite{shuman2013emerging}.

This allows us to define a graph signal, denoted by the vector $\bsx \in \mathbb{R}^{N}$, as a  mapping from the node set to the set of real vectors; that is, $\bsx : \ccalV \to \mathbb{R}^{N} $. In this way, $x_i \in \mathbb{R}$ is a scalar that represents the signal value at node $i$. Because $\bbS$ reflects the local connectivity of $\ccalG$, the operation $\bbS\bsx$ performs at each node a local computation enabling us to introduce the concept of filtering in the graph setting.

\textbf{Graph Filters}
We can process a graph signal $\bsx$ by means of a so-called graph filter \cite{6409473} as:
\begin{equation}
\label{eq:filtering}
        \bsy= \bbH \!\left( \bsh, \bbS\right) \bsx= \sum_{k=0}^{K}  h_{k}   \bbS^{k} \bsx,
\end{equation}
where $K$ is the order of the filter, $\bbH \!\left(\bsh, \bbS\right)$ is a polynomial matrix on $\bbS$ and $\bsh:= [ h_0, \ldots, h_K]^{\top}$ is the vector that contains the filter taps. Due the locality of $\bbS$, graph filters represent linear transformations that can be implemented in a distributed setting \cite{segarra2017network}. More formally, the output entry $y_i$ of $\bsy$ at node $i$ is a linear combination of $K+1$ terms: the first term is the signal value $x_i$ of node $i$; the $k$th term ($k=1,2,\dots,K$) combines signal values $x_j$ from  the $k$-hop neighbors of node $i$.

\textbf{Topology Identification}
When the connections of the network cannot be directly observed or the network is just a conceptual model of pair-wise relationships among entities, a fundamental question is how to learn its structure from the graph signals. Formally, consider the matrix $\bbX=\left[\bsx_1, \ldots, \bsx_T\right] \in \mathbb{R}^{N \times T}$ that stacks column-wise $T$ graph signals $\bsx_t$ residing over the network $\ccalG=\left(\ccalV,\ccalE, \bbS \right)$. The goal is to infer the \textit{latent} underlying network topology encoded in the GSO $\bbS$ under some optimality criterion. 

This problem has been addressed in the past by means of statistical approaches, mostly based on correlation analysis and its connections to covariance selection and high-dimensional regression for learning Gaussian graphical models. Only more recently, GSP postulated the network topology inference problem under the assumption that the observed signals exhibit certain properties over the graph, such as smoothness, stationarity or band-limitedness. The reader interested in this topic is referred to \cite{mateos2019connecting} \cite{dong2019learning} \cite{giannakis2018topology}.

Differently from the traditional topology identification setting, instead of estimating ${\bf S}$ from ${\bf X}$, we rely on model~\eqref{eq:filtering} and focus on a problem where given input and output data, the values of the nonzero entries of ${\bf S}$, i.e., the edge  weights, and the filter taps $\bsh$ of a graph filter ${\bf H}(\bsh, {\bf S})$ have to be jointly identified.
In Section~\ref{sec:joint}, we rigorously formulate this problem and, in Section~\ref{sec:algorithm}, we propose a way to efficiently tackle it.

\section{Joint graph filter and topology estimation}
\label{sec:joint}

Suppose there is an unknown network process that can be accurately modelled by a graph filter $\bbH \! \left(\bsh, \bbS\right)$ where, in response to an input $\bsx_t$, we observe a corresponding output $\bsy_t$.  Such dynamics can be found for instance in social networks, where as a result of an advertisement campaign, we may expect to observe a response of the network's users; or in epidemics, where the nodes of the network are cities and we monitor the evolution of a spreading disease from one time instant to the next.

Let us assume that there are $T$  input-output pairs available, and that we stack them column-wise in the matrices $\bbX=\left[\bsx_1, \ldots, \bsx_T\right]$ and $\bbY=\left[\bsy_1, \ldots, \bsy_T\right]$, respectively.
Let the unknown filter $\bbH \! \left(\bsh, \bbS\right)$ be of the form in \eqref{eq:filtering}. At this point, we are ready to formally state the problem we are going to address.

\noindent \textbf{Problem Statement} \textit{Given the input-output data $\{\bsx_t, \bsy_t \}_{t=1}^{T}$ and the support, $\mathcal{A}$, of the graph $\ccalG$, the goal is to identify the filter coefficients $\bsh$ and the GSO $\bbS$ embodied in the graph filter  $\bbH \! \left(\bsh, \bbS \right)$, that maps $\bsx_t$ into $\bsy_t$ as accurately as possible.}

The above problem can be mathematically defined with a least-squares formulation as:  
\begin{equation}
\label{eq:LS-cons-final}
\begin{array}{cl}
\underset{\bsh, \bbS}{\argmin}
&  \|  \bbY - \sum_{k=0}^{K}  h_{k}   \bbS^{k} \bbX\|_{\rm F}^{2} \\
\text{s.t.} & \bbS \in \ccalS \\
& \operatorname{supp}\left(\bbS\right) \subseteq \mathcal{A}
\end{array}
\end{equation}
where $\ccalS$ represents the set of valid GSOs, 
$\mathcal{A}$ denotes the set with the support of $\mathcal{G}$,
and $\Vert \cdot \Vert_{\rm F}$ denotes the Frobenius matrix norm. Note the (relaxed) constraint on the support: as the sparsity pattern of the GSO might have been overestimated, we leave it to the algorithm to optimize it, eventually shrinking to zero some unnecessary edges. That is, we constrain only the entries of the GSO to be zero in correspondence to the zeros of the support, leaving the other entries unconstrained (both zero and non-zero values are admitted).

From~\eqref{eq:LS-cons-final}, we can deduce that the problem is not convex. Indeed, the objective function is made up of cross-products between the entries of $\bbS$ and the filter coefficients $h_{k}$, and by the power terms $\bbS^{k}$. The overall optimization problem is hence not convex and traditional tools of convex optimization cannot be used. 

Although not directly handling the fixed-support case, the works referenced in Section~\ref{sec:intro} address the estimation problem using multi-step approaches to find $\bbS$ and/or $\bsh$.  For instance, in \cite{7763882} each realization is modeled through a graph filter-based vector auto-regressive (VAR) model, and this structure is leveraged to first recover the graph filters $\bbH_i(\bsh, \bbS)$  representing the matrix filter taps of the VAR, and only then to recover the shift $\bbS$ and the coefficients $\bsh$ from them. Other approaches, such as \cite{segarra2017network} \cite{shafipour2018identifying} are only interested in learning the shift $\bbS$, while others, such as \cite{segarra2016blind}, only in the filter coefficients $\bsh$.

Differently from the method in \cite{7763882}, in the following, we introduce a globally convergent SCP-based method to directly find both the filter taps $\bsh$ and the GSO $\bbS$. 
To the best of our knowledge, this is the first work that
jointly learns the filter taps and the graph topology from observations. 

\section{Alternating Minimization}
\label{sec:algorithm}

To tackle the non-convexity of the problem and to bypass the limited flexibility of other methods, we resort to the alternating minimization (AM) approach, acting iteratively on $\bsh$ and $\bbS$.
The general AM pseudo-code, adapted to our case, is reported in Algorithm~\ref{alg:complete}. Notice that due to steps $3$ and $4$ in Algorithm~\ref{alg:complete}, the cost is guaranteed to be a non-increasing function of the iteration number. In the following, we show how to perform step $3$ and $4$ of the proposed algorithm.

\begin{algorithm}[t]
\begin{algorithmic}[1]
\Require Feasible $\bbS^{\left(0\right)}$, $\varepsilon>0$, $\mathcal{A}$, $\mathcal{S}$
\State $n \leftarrow 1$
\While{not converged}

\State $\bsh^{\left(n\right)} \leftarrow \argmin_{\bbh} f\left(\bsh, \bbS^{\left(n-1\right)} \right)$ \quad \quad \quad [See Eq.~\eqref{eq:h-learning}]

\State $\bbS^{\left(n\right)} \leftarrow \argmin_{\bbS} f\left(\bsh^{\left(n\right)}, \bbS \right)$ \quad \quad \quad (\textbf{SCP}) [See Alg.~\ref{alg:complete2}]

\State Check convergence ($\bsh^{\left(n\right)}, \bbS^{\left(n\right)}, \varepsilon$)
\State $n \leftarrow n + 1$
\EndWhile
\State \Return $\bbS^{\left(n\right)}, \bsh^{\left(n\right)}$
\end{algorithmic}
\caption{Joint GF \& GSO Identification}
\label{alg:complete}
\end{algorithm}

Given the estimate of the GSO $\bbS$ at the $(n-1)$th iteration, i.e., $\bbS^{(n-1)}$, the estimation problem at the $n$th iteration for the filter taps vector $\bsh$, i.e., $\bsh^{(n)}$, reads as:
\begin{equation}
    \label{eq:h-learning}
    \begin{aligned}
        \bsh^{(n)} = & &  \underset{\bsh}{\argmin}
        &   \|  \bbY - \sum_{k=0}^{K}  h_{k}   \big(\bbS^{(n-1)}\big)^{k} \bbX\|_{\rm F}^{2}.\\
    \end{aligned}
\end{equation}
 Problem \eqref{eq:h-learning} is convex and boils down to the traditional linear least squares (LLS) problem. 
 
 The solution of \eqref{eq:h-learning} is then used in the next step, i.e., step $4$, to  minimize the function with respect to the constrained GSO $\bbS$; that is 
 \begin{equation}
    \label{eq:S-learning}
        \begin{array}{cccl}
        \bbS^{\left(n\right)} &=& \underset{\bbS}{\argmin}&  \{f(\bbS) := \| \bbY - \sum_{k=0}^{K}  h_{k}^{\left(n\right)}   \bbS^{k} \bbX\|_{\rm F}^{2}\} \\
         &&\text{s.t.} & \bbS \in \ccalS \\
         &&& \operatorname{supp}\left(\bbS\right) \subseteq \mathcal{A} 
    \end{array}
\end{equation}
As problem~\eqref{eq:S-learning} is not convex, we employ SCP~\cite{boyd2008sequential}, a heuristic and local optimization method for non-convex problems that leverages convex optimization, where the non-convex portion of the problem is modeled by convex functions that are (at least locally) accurate. 

Given the non-convex function $f(\bbS)$, the idea in SCP is to maintain a solution estimate $\bbS^{[l]}$ and a respective convex \textit{trust region} $\mathcal{T}^{[l]} \subseteq \mathbb{R}^{N \times N}$ over which we ``trust'' our solution to reside\footnote{We use the superscript with square brackets to indicate the SCP iterations.}. Then, using a convex approximation $\widehat{f}$ of $f$, around $\bbS^{[l]}$, the next solution estimate, $\bbS^{[l+1]}$, is computed using the optimizer of $\widehat{f}$ in $\mathcal{T}^{[l]}$. Typical trust regions include $\ell_2$-norm balls or bounded regions. 

For our case, we define as trust region the box:
\begin{equation}\label{eq.Teq}
    \mathcal{T}^{[l]}=
    \begin{cases} 
    \bbS \in \ccalS,  \\
      \operatorname{supp}\left(\bbS\right) \subseteq \mathcal{A} \\ 
     |[\bbS]_{ij} -[\bbS^{[l]}]_{ij} | \leq \rho_{ij}(l), \: \text{if } (i,j)\in\mathcal{E} \neq 0, \:  \forall i,j \in \ccalV,
    \end{cases}
\end{equation}
where $\rho_{ij} : \mathbb{Z}_+ \rightarrow \mathbb{R}_{++}$ is a mapping from the iteration number to the breadth of the search for the $(i,j)$th entry.

For the convex approximation of the function, we linearize the function $f(\bbS)$ around the previous estimate $\bbS^{[l]}$ using its first-order Taylor approximation\footnote{The computation of $\nabla_{\bbS}f(\bbS^{[l]})$ is reported in the Appendix.}:
\begin{equation} \label{eq.fhat}
\hat{f}^{[l]}({\bf S})  := f({\bf S}^{[l]}) + \operatorname{tr}\left[\nabla_{\bbS}f(\bbS^{[l]})^\top (\bbS - \bbS^{[l]})\right].
\end{equation}
We then find a feasible intermediate iterate by solving the problem:
\begin{equation}
\label{eq:S_hat}
    \hat{\bf S} = \underset{{\bf S}\in\mathcal{T}^{[l]}}{\arg\min} \; \hat{f}^{[l]}({\bf S}).
\end{equation}

Due to the non-convexity of the cost function $f(\bbS)$, its value at the (feasible) point $\hat{\bf S}$ is not guaranteed to be lower than the one at $\bbS^{[l]}$. Hence, to find the ``best'' feasible solution $\bbS$ at the $(l+1)$th iteration, we first resort to a line search to find the optimal scaling step size parameter $\alpha_l$ toward the feasible descent direction $\Delta_l := \hat{\bf S} - {\bf S}^{[l]}$; that is,
\begin{equation}
\label{eq:alpha_opt}
    \alpha_l^* = {\arg\min}_{\alpha_l\in[0,1]} f({\bf S}^{[l]} + \alpha_l\Delta_l).
\end{equation}
Then, we compute our next solution estimate $ {\bf S}^{[l+1]}$ through
\begin{equation}
\label{eq:S_new}
    {\bf S}^{[l+1]} = {\bf S}^{[l]} + \alpha_l^{*}\Delta_l,
\end{equation}
which is feasible for the original problem as long as the set $\mathcal{S}$ is convex, i.e., the update in~\eqref{eq:S_new} is a convex combination of feasible points. 
The specialized SCP procedure for our problem is summarized in Algorithm~2. Note that steps $7$-$9$ guarantee, at each iteration, the feasibility of the iterate and a non-increasing cost function value, leading to the global convergence of Algorithm~\ref{alg:complete}.

\begin{algorithm}[]
\begin{algorithmic}[1]
\Require $\bbS^{\left(n\right)}$, $\bsh^{(n)}$, $\{\rho_{ij}\}_{(i,j)\in\mathcal{E}} $ , $\varepsilon>0$
\State $l \leftarrow 1$
\State $\bbS^{[0]} \leftarrow \bbS^{\left(n\right)}$
\While{not converged}
\State Compute  $\{\rho_{ij}(l-1)\}_{(i,j)\in\mathcal{E}}$
\State Construct $\hat{f}^{[l-1]}({\bf S})$ as in \eqref{eq.fhat}
\State Define $\mathcal{T}^{[l-1]}$ as in~\eqref{eq.Teq}
\State $\hat{\bf S} \leftarrow \underset{{\bf S}\in\mathcal{T}^{[l-1]}}{\arg\min} \; \hat{f}^{[l-1]}({\bf S})$ 
\State $ \alpha_{l-1}^* \leftarrow {\arg\min}_{\alpha_{l-1}\in[0,1]} f({\bf S}^{[l-1]} + \alpha_{l-1}(\hat{\bf S} - {\bf S}^{[l-1]}))$
\State ${\bf S}^{[l]} \leftarrow {\bf S}^{[l-1]} + \alpha_{l-1}^{*}(\hat{\bf S} - {\bf S}^{[l-1]}))$
\State Check convergence ($\bsh^{\left(n\right)}, \bbS^{[l]},\varepsilon$)
\State $l \leftarrow l + 1$
\EndWhile
\State \Return $\bbS^{[l]}$
\end{algorithmic}
\caption{SCP}
\label{alg:complete2}
\end{algorithm}

Due to the non-convexity of the cost function,  the global optimality of the solution is not guaranteed, thus the results are dependent on the initial starting point(s) as they might lead to different local minima. Despite that in these cases multi-start is recommended, we have found in our numerical experiments that both the unweighted adjacency matrix, $\bf A$, and the respective combinatorial Laplacian matrix, $\bf L$, are good initial iterates, i.e., ${\bf S}^{(0)}$, for the proposed approach; they are straightforward choices and can be computed using the support of the graph. To validate this claim, in our experiments, we generate initial GSO iterates $\bbS_i^{\left(0\right)}$, through a method reported in the Appendix, and show their performance in the next section, along with those of $\bf A$ and $\bf L$.


\begin{figure*}%
\centering
\begin{subfigure}{0.33\textwidth}
\centering
\psfrag{NMSE}[bc]{\footnotesize NMSE}
\psfrag{Iteration Number}[cl]{\footnotesize Iteration Number}
\psfrag{S1}{\tiny{$\bbS_\!1$}}
\psfrag{S2}{\tiny$\bbS_\!2$}
\psfrag{S3}{\tiny$\bbS_\!3$}
\psfrag{S4}{\tiny$\bbS_\!4$}
\psfrag{S5}{\tiny$\bbS_\!5$}
\psfrag{BL}{\tiny${\bf L}$}
\psfrag{BA}{\tiny${\bf A}$}
\includegraphics[width=.95\textwidth]{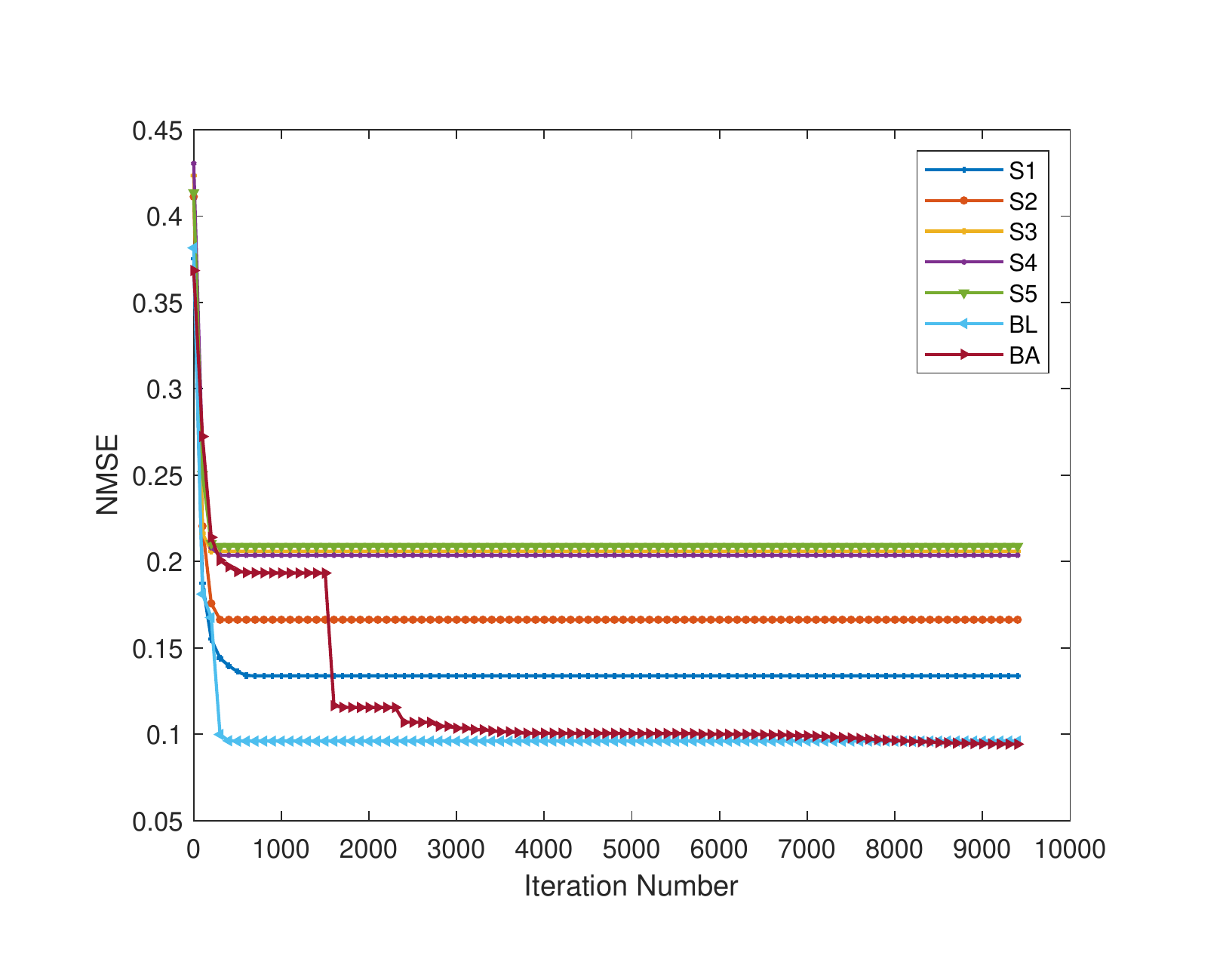}%
\caption{$\text{S}_{g}=$ L, $\text{S}_{h}=$ L}%
\label{sfig_lapllapl}%
\end{subfigure}%
\begin{subfigure}{0.32\textwidth}
\centering
\psfrag{NMSE}[bc]{\footnotesize NMSE}
\psfrag{Iteration Number}[cl]{\footnotesize Iteration Number}
\psfrag{S1}{\tiny{$\bbS_\!1$}}
\psfrag{S2}{\tiny$\bbS_\!2$}
\psfrag{S3}{\tiny$\bbS_\!3$}
\psfrag{S4}{\tiny$\bbS_\!4$}
\psfrag{S5}{\tiny$\bbS_\!5$}
\psfrag{BL}{\tiny${\bf L}$}
\psfrag{BA}{\tiny${\bf A}$}
\includegraphics[width=.95\textwidth]{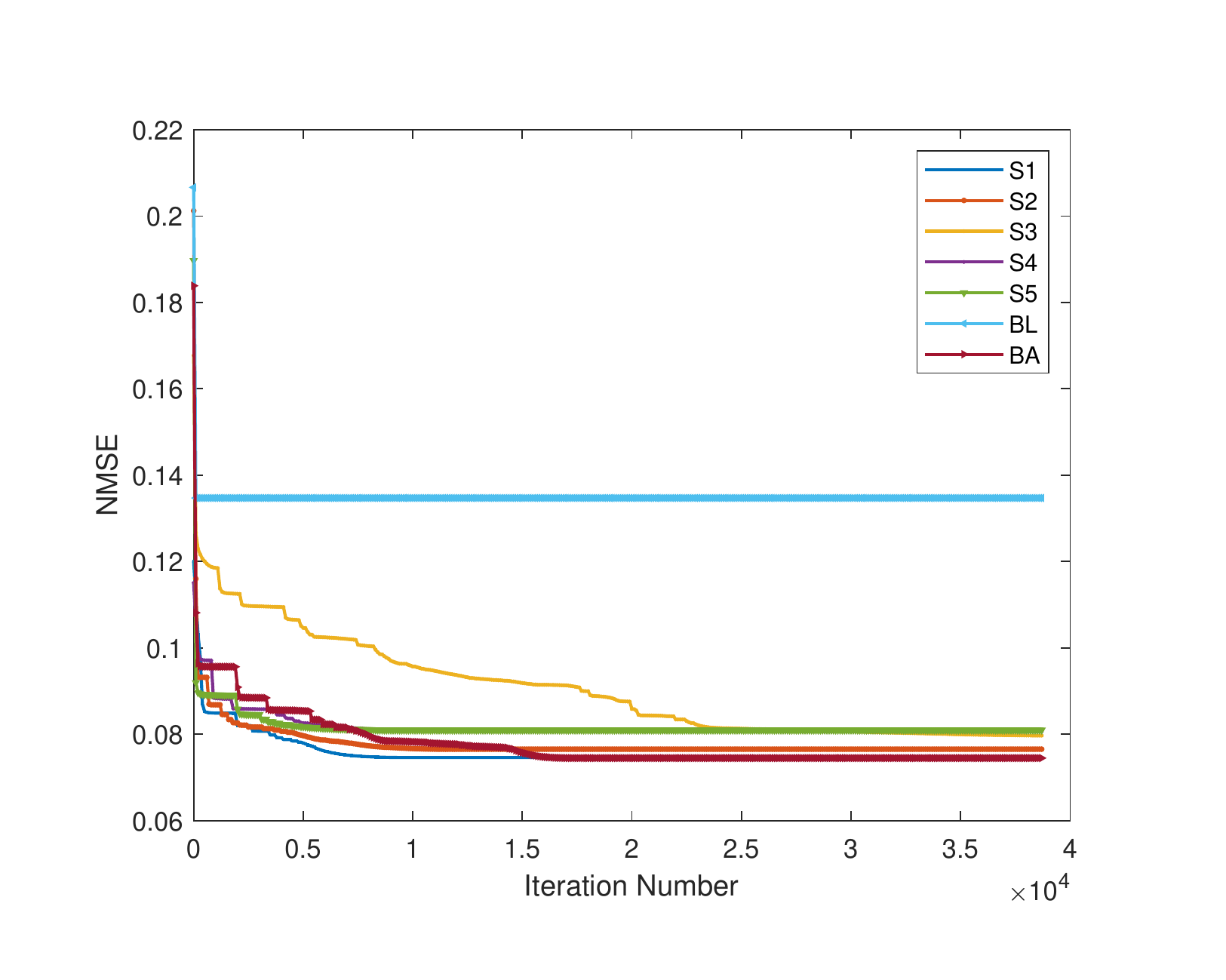}%
\caption{$\text{S}_{g}=$ W, $\text{S}_{h}=$ W}%
\label{sfig_adjadj}%
\end{subfigure}%
\begin{subfigure}{0.32\textwidth}
\psfrag{NMSE}[bc]{\footnotesize NMSE}
\psfrag{Iteration Number}[cl]{\footnotesize Iteration Number}
\psfrag{S1}{\tiny{$\bbS_\!1$}}
\psfrag{S2}{\tiny$\bbS_\!2$}
\psfrag{S3}{\tiny$\bbS_\!3$}
\psfrag{S4}{\tiny$\bbS_\!4$}
\psfrag{S5}{\tiny$\bbS_\!5$}
\psfrag{BL}{\tiny${\bf L}$}
\psfrag{BA}{\tiny${\bf A}$}
\centering
\includegraphics[width=.95\textwidth]{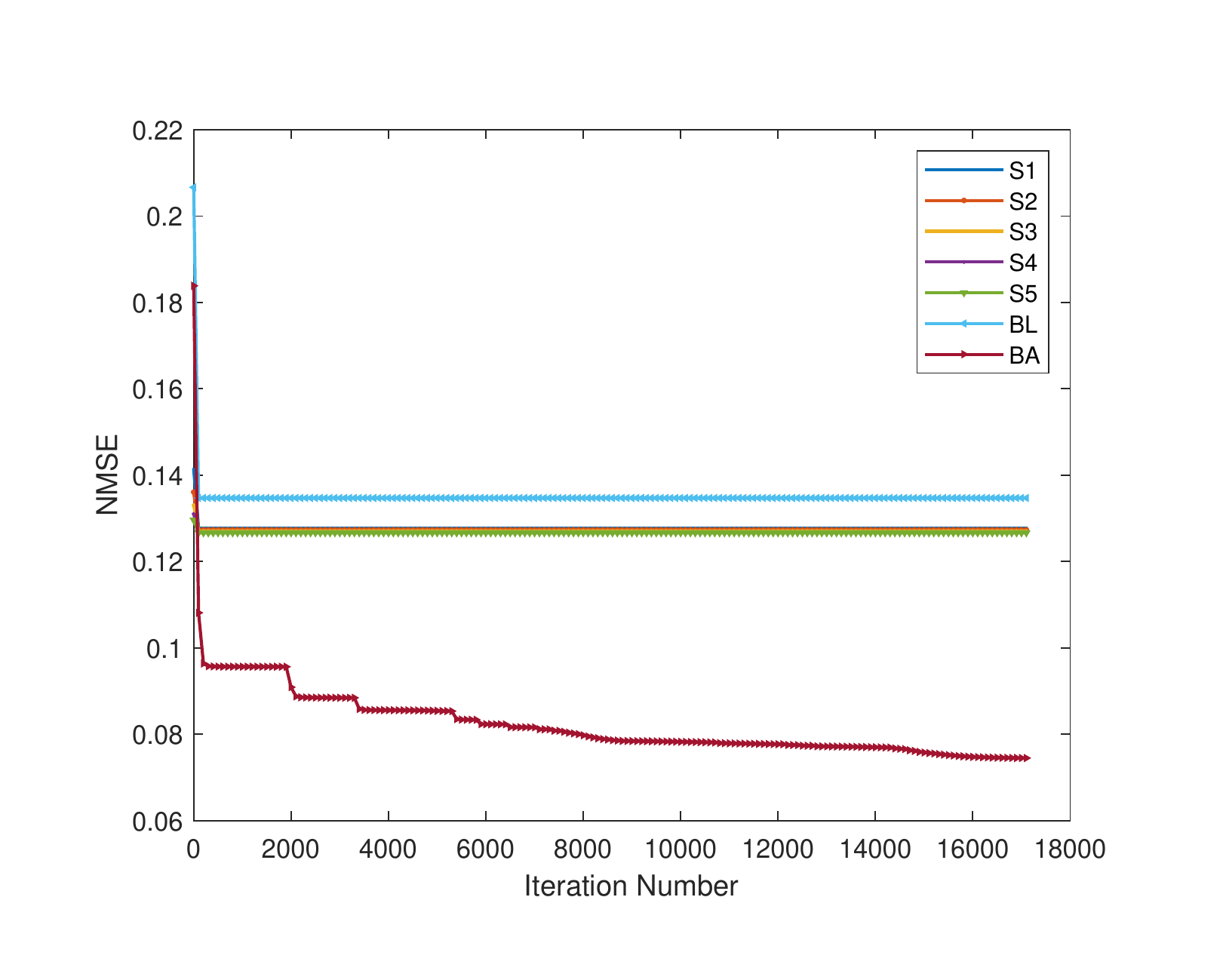}%
\caption{$\text{S}_{g}=$ W, $\text{S}_{h}=$ L}%
\label{sfig_adjlapl}%
\end{subfigure}%
\caption{NMSE for different settings of the true  GSO type $\text{S}_{g}$ and the hypothesis GSO type $\text{S}_{h}$. The legend in each plot contains the considered GSOs for initializing the algorithm.}
\label{fig:vary_n1}\vspace{-5mm}
\end{figure*}


\section{Numerical Results}
\label{sec:num-res}
In this section, we show some numerical results obtained for identifying different graph filters and GSOs $\bbS$. In these experiments, we consider cases where the GSOs to identify are the weighted adjacency matrix and the Laplacian.


To evaluate the correctness of our method, we first generate a random graph composed of $N=30$ nodes with the GSP Toolbox \cite{perraudin2014gspbox} and construct from the graph the respective GSO $\bbS$ involved in the graph filter that generates the output data. 
We then generate $T=500$ input graph signals $\{\bsx_t\}_{t=1}^{T}$ drawn from a standard normal distribution. 
By fixing the order of the graph filter to $K=5$, we generate graph filter taps $\bsh$ following a Gaussian distribution with zero mean and $\sigma = 3$. Finally, the output graph signals $\{\bsy_t\}_{t=1}^{T}$ are generated following~\eqref{eq:filtering}.

In our experiments, we analyze two main aspects of the proposed method: \textit{i)} the convergence of the algorithm, regardless the initial starting point; and \textit{ii)} the similarity in terms of edge weights between the groundtruth GSO $\bbS$ and the identified one $\widehat{\bbS}$. To provide a fair comparison, we assume we do not know in advance the type of GSO that generate the network process, i.e., $\mathcal{S}$ is not completely known a priori. For this, we provide a guess of GSO type as input to Algorithm~1, and hope that a proper guess leads to a good fitting. In the sequel, we denote with $\text{S}_{g}$ the type of GSO used to \textit{generate} the data, and with $\text{S}_{h}$ the type of GSO \textit{hypothesized}. Both  types of GSOs can assume the values W and L, indicating respectively the (weighted) adjacency matrix and the Laplacian matrix\footnote{Note how we don't use here the bold notation, because both $\text{S}_g$ and $\text{S}_h$ are (textual) parameters of the algorithm, in contrast to the considered GSOs starting points that are effectively matrices.}.

As performance metric for the error evaluation, we consider the normalized MSE (NMSE), defined as

\begin{equation}
    \label{eq:nmse}
    \text{NMSE}= \frac{\sum_{t=1}^{T}\left\|\widehat{\mathbf{y}}_{t}-\mathbf{y}_{t}\right\|_{2}^{2}}{\sum_{t=1}^{T}\left\|\mathbf{y}_{t}\right\|_{2}^{2}}
\end{equation}
where $\widehat{\mathbf{y}}_{t}$ is the predicted graph signal relative to the input  $\bbx_t$.

Figure~\ref{sfig_lapllapl} shows the NMSE as function of the ``cumulative'' iteration number\footnote{We count all the iterations of the algorithm up to its convergence. We sum in a cumulative manner the outer and the inner iterations of Algorithm~1.}, for $\text{S}_g=\text{S}_h= $ L. Regardless of the starting point, we observe the non-increasing behavior of the NMSE, corroborating the global convergence of the algorithm. For this particular $(\text{S}_g,\text{S}_h )$ combination, $\bf L$ and $\bf A$ are the best performing starting points in terms of final NMSE, with $\bf L$ reaching convergence in just a few iterations. The sharp  steps downwards, especially noticeable in the case of $\bf A$ are due to the update of the graph filter coefficients $\bsh$. In this case, the other initial points are not better that the straightforward initial guesses. 
Similar observations can be made from Fig.~\ref{sfig_adjadj}.
A case of GSO mismatch is shown in Fig.~\ref{sfig_adjlapl}, where the data are generated using the weighted adjacency matrix, but the algorithm is running based on  the Laplacian hypothesis. As expected, the $\bf A$ matrix is the best starting point. Comparing Fig.~\ref{sfig_adjadj} and  Fig.~\ref{sfig_adjlapl}, where the curves starting at $\bf A$ and $\bf L$ achieve the same NSMSE, we note how in case of matched hypotheses, the GSOs generated through the generation procedure yield a lower error with respect to the mismatched counterpart.
 
\begin{figure}[h!]
     \centering
     \includegraphics[width=0.7\columnwidth, trim= 1cm 0.5cm 0cm 0cm, clip=false ]{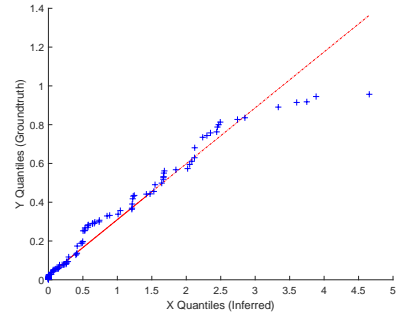}
     \caption{Q-Q plot of the weights of the groundtruth GSO and the weights of the inferred Laplacian for the case $\text{S}_{g}=$ L, $\text{S}_{h}=$ L }
     \label{fig:qqplot}
 \end{figure}

\begin{figure}[h]
\centering
\begin{subfigure}{\columnwidth}
\centering
\includegraphics[width=\textwidth, trim=  1cm 0.5cm 1cm 0.5cm, clip=true]{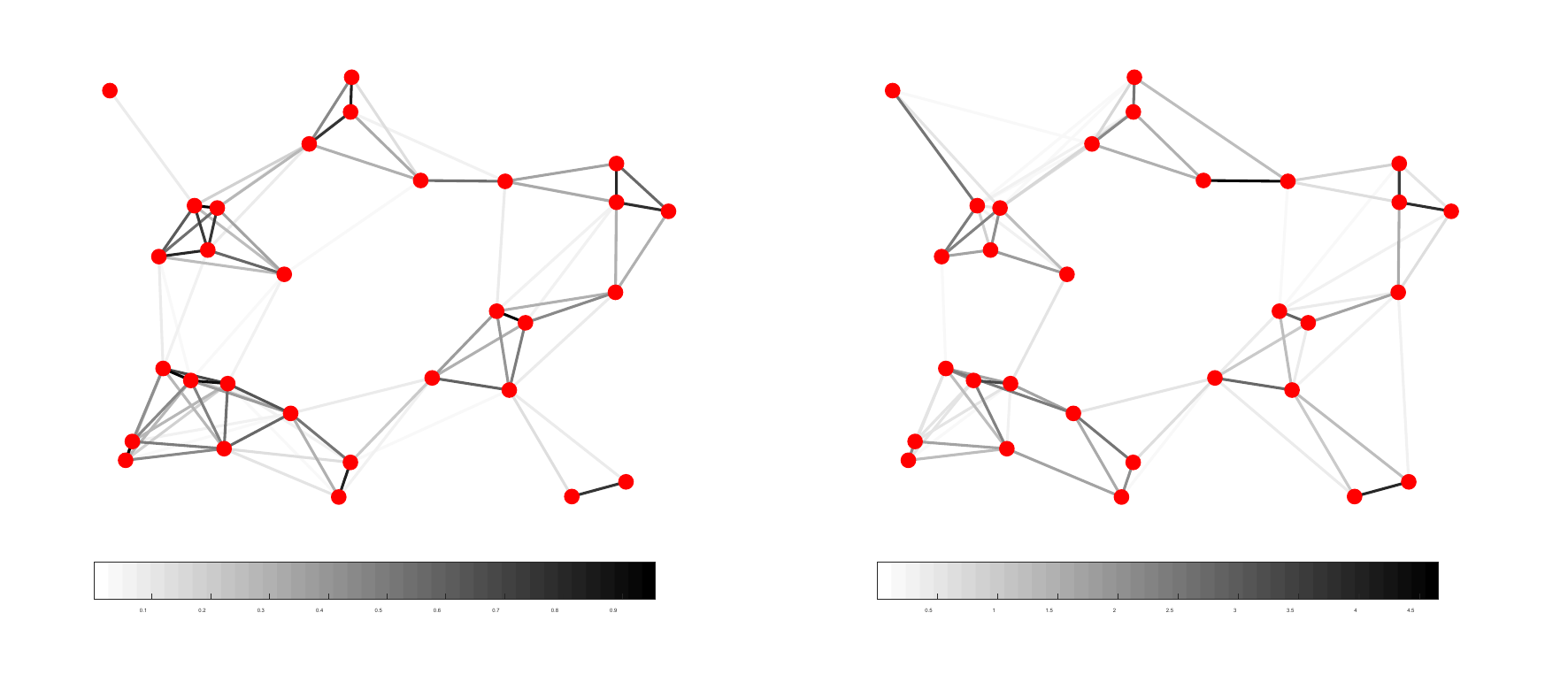}%
\caption{Left: groundtruth graph. Right: inferred graph. The initial condition for the inferred graph was L. The darker the edge in the graph, the higher its value.}%
\label{sfig_graphs}%
\end{subfigure}%
\newline
\begin{subfigure}{\columnwidth}
\centering
\includegraphics[width=\textwidth, trim= 2cm 1.5cm 1cm 1cm, clip=true]{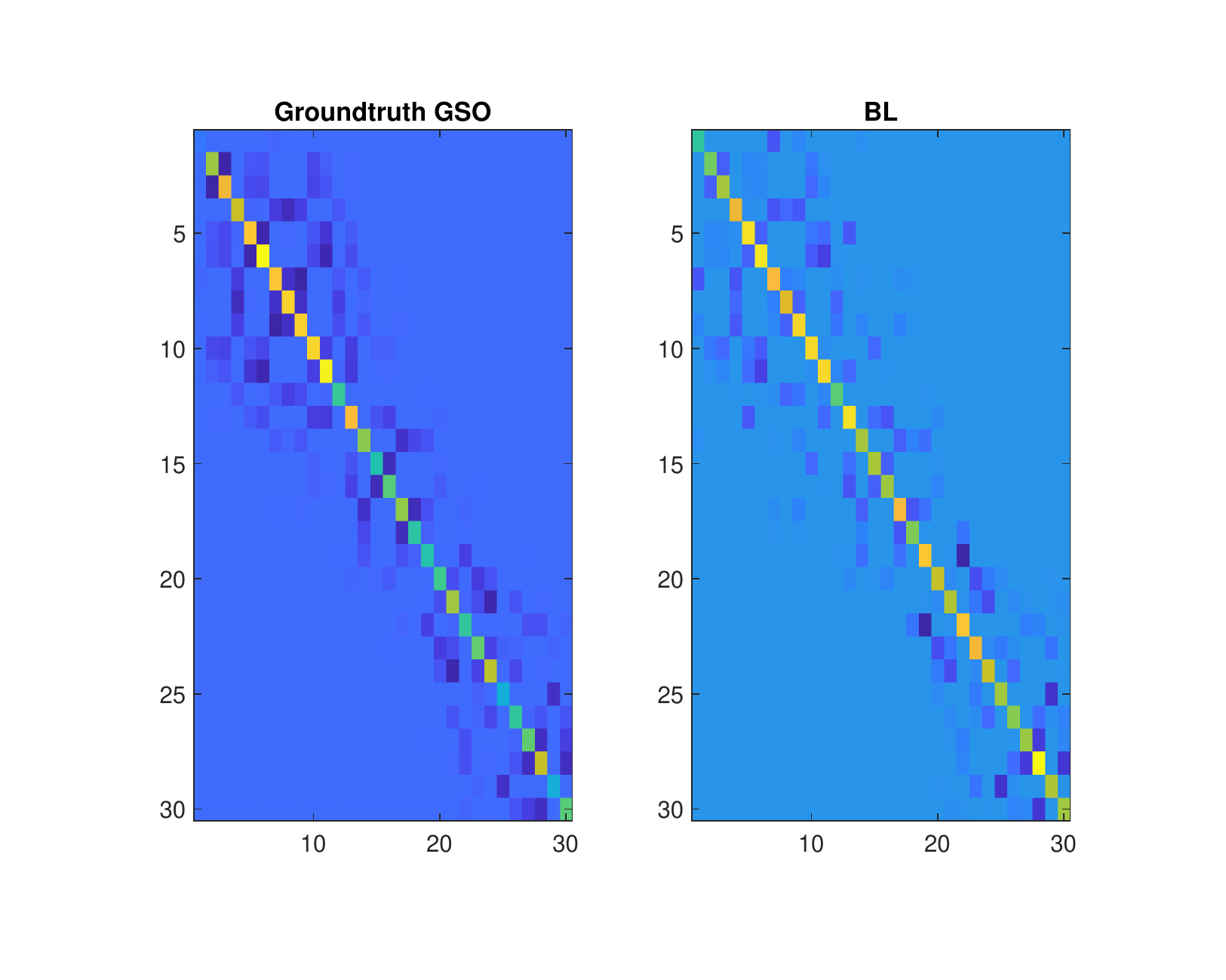}%
\caption{Left: groundtruth Laplacian. Right: inferred Laplacian}%
\label{sfig_heatmaps}%
\end{subfigure}%
\caption{ (a) Graphs and (b) Laplacian matrix heatmaps for the case $\text{S}_{g}=\text{S}_{h}=$L}\label{fig:qualitative}
\end{figure}

 As a quantitative measure of similarity between the groundtruth and the inferred weights, we report their Spearman correlation coefficient $r_s$, which is a non parametric measure of rank correlation. In particular, it answers the following question: \emph{do edges with higher weight in the groundtruth GSO tend to have a higher weight in the inferred one?} A perfect Spearman correlation of $+1$ or $-1$ occurs when each of the variables is a perfect monotone function of the other. In our setting, $r_s=0.74$ thus confirming a strong positive correlation of the two vectors.  
 Moreover, as depicted in the Q-Q plot of Fig.~\ref{fig:qqplot}, the quantiles of the two vectors lie almost entirely on the straight line, allowing us to state that the weights of the two GSOs come approximately from the same distribution.
 For a qualitative and visual assessment of the method, in Fig.~\ref{sfig_graphs} and Fig.~\ref{sfig_heatmaps} we show, respectively, the graphs and the weighted sparsity pattern of the groundtruth and the learned Laplacian matrix (for $\text{S}_{g}=\text{S}_{h}=$ L).  
We observe how, up to a scaling factor, the algorithm is able to give a larger weight to those edges that are  also ``important'' in the original graph. All these considerations make us optimistic in the continuation of the development and the study of the proposed approach, driving us toward its application in more complex real-world scenarios.

 \section{Conclusion}
 In this work, we formulated and studied the problem of jointly estimating the filter coefficients and the graph shift operator (GSO) defining a graph filter that models the dynamics of signals defined over a network. In particular, motivated by practical scenarios, we exploited the a priori knowledge of the  sparsity pattern of the network. We proposed an alternating-minimization approach, whose non-convex subproblem is handled through sequential convex programming methods. As shown in the numerical results, the proposed method is globally convergent and is able to identify the type of GSO used to generate the data. Quantitative statistical measures and qualitative graphics demonstrated the efficacy of the algorithm to assign higher values to those weights that are prominent in the real graph. 

\section{Appendix}
\label{sec: appendix}
To compute the derivative $\nabla_{\bbS}f(\bbS)$, let us first expand the function $f(\bbS)$\footnote{We set $\bsh^{(n)}$ to $\bsh$, and $\bbH(\bsh, \bbS)$ to $\bbH$, for the rest of the proof.}:
\begin{align*}
f(\bbS) &=\\
&= \tr \left[(\bbY - \bbH(\bsh, \bbS)\bbX)(\bbY - \bbH(\bsh, \bbS)\bbX)^{\top} \right] \\
&= \tr\left(\bbY\bbY^{\top}\right) - 2 \tr\left(\bbH\bbX\bbY^{\top}\right) + \tr\left[ \bbH^{\top}\bbH\bbX\bbX^{\top}\right] \\
&=\tr\left[\bbY\bbY^{\top}\right] - 2  \sum_{k=0}^{K} h_{k}\tr\left[ \bbS^{k}\bbX \bbY^{\top} \right] \\
& \quad\quad+ \sum_{k_{1}}^{K} \sum_{k_{2}}^{K} h_{k_{1}} h_{k_{2}} \tr\left(\bbS^{k_{1} + k_{2}} \bbX\bbX^{\top}\right)
\end{align*}

Then 
\begin{align*}
\nabla_{\bbS}f(\bbS) &= \\
&= - 2  \sum_{k=0}^{K} h_{k} \nabla_{\bbS} \tr\left[ \bbS^{k} \bbX \bbY^{\top} \right] \\
& \quad \quad + \sum_{k_{1}}^{K} \sum_{k_{2}}^{K} h_{k_{1}} h_{k_{2}} \nabla_{\bbS}\tr\left(\bbS^{k_{1} + k_{2}} \bbX\bbX^{\top}\right)
\end{align*}

Because $\bbS$ is symmetric, we have to take into account its structure for the computation of the derivative of $f(\bbS)$. Indeed, due to the matrix symmetry, the overall gradient can be decomposed in:

\begin{equation*}
    \nabla_{\bbS}f(\bbS)=\left[\frac{\partial f(\bbS)}{\partial \mathbf{S}}\right]+\left[\frac{\partial f(\bbS)}{\partial \mathbf{S}}\right]^{\top}-\operatorname{diag}\left[\frac{\partial f(\bbS)}{\partial \mathbf{S}}\right].
\end{equation*}
Finally, because  $\frac{\partial}{\partial \mathbf{S}} \operatorname{Tr}\left(\mathbf{S}^{k}\right)=k\left(\mathbf{S}^{k-1}\right)^{\top}$ and $\frac{\partial}{\partial \mathbf{S}} \operatorname{Tr}\left(\mathbf{BS}^{k}\right)= \sum_{r=0}^{k-1}\left(\mathbf{S}^{r} \bbB \mathbf{S}^{k-r-1}\right)^{\top}$, we have that the component $\left[\partial f(\bbS) / \partial \mathbf{S}\right]$ of the gradient is :
\begin{align*}
\frac{\partial f(\bbS)}{\partial \mathbf{S}} &=  \\
&= - 2  \sum_{k=0}^{K} h_{k} \nabla_{\bbS} \tr\left[ \bbS^{k} \bbX \bbY^{\top} \right] \\
& \quad \quad + \sum_{k_{1}}^{K} \sum_{k_{2}}^{K} h_{k_{1}} h_{k_{2}} \nabla_{\bbS}\tr\left(\bbS^{k_{1} + k_{2}} \bbX\bbX^{\top}\right) \\
&= - 2  \sum_{k=1}^{K} h_{k} \left[\sum_{r=0}^{k-1}  (\bbS^{r} \bbX\bbY^{\top} \bbS^{k-r-1})^{\top} \right] \\
& \quad \quad + \sum_{k_{1}}^{K} \sum_{k_{2}}^{K} h_{k_{1}} h_{k_{2}} \sum_{r=0}^{k_{1}+k_{2}-1}  (\bbS^{r} \bbX\bbX^{\top} \bbS^{k_{1}+k_{2}-r-1})^{\top}
\end{align*}

 \subsection{GSO Candidate Generation}
\label{sec:candidate}
Let the model be $\bsy=\bbH(\bsh, \bbS)\bsx$ for some order $K$ of the filter. 
 Then, a $K=1$ approximation for the overall problem \eqref{eq:LS-cons-final} is given by
 \begin{align}
  \label{eq:approx1}
     \bsy \approx (\hat{h}_0^{(1)}\bbI + \hat{\bbS}_{1})\bsx,
 \end{align}
 where $\hat{h}_0^{\left(1\right)}$ is the constant filter tap estimate related to the first order approximation, and $\hat{\bbS}_{1} \in \ccalS$ is the respective estimate for the GSO. We assume $\hat{h}_1^{(1)}=1$ to avoid the scalar ambiguity that would otherwise arise in the term $\hat{h}_1 \hat{\bbS}_1$. This also decouples the filter parameters from the GSO, both of which can be estimated by LLS. This way, a first GSO candidate $\bbS_1^{\left(0\right)}$ for the algorithm is found. 
 Next, we consider a second order approximation of the model

 \begin{align}
\label{eq:approx2}
    \bsy \approx (\hat{h}_0^{(2)}\bbI + \hat{\bbS}_{2} +  \hat{h}_2^{(2)}\bbS_{1}^{\left(0\right)2})\bsx,
\end{align}
 where the variables are now $\hat{h}_0^{(2)}, \hat{h}_2^{(2)} $ and $\hat{\bbS}_{2}$, and we still assume the first filter tap $\hat{h}_1^{(2)}$ is equal to one. This again leads to a LLS problem which generates a second GSO candidate $\bbS_2^{\left(0\right)}$. 
 We iterate this procedure by increasing the order of the filter at each step, and maintaining the term that is linear in the GSO variable $\bbS$. At the end, we have $K$ initial GSO candidates $\bbS_1^{\left(0\right)}, \bbS_2^{\left(0\right)}, \ldots, \bbS_K^{\left(0\right)}$, which can be given as input to Algorithm 1.


\bibliographystyle{IEEEbib}
\bibliography{refs}
\end{document}